\documentclass{appolb}
\usepackage{epsfig}

\def\q{\mbox{\boldmath $q$}}
\usepackage[T1]{fontenc}
\begin{document}
\eqsec  
\title{Quasielastic Charged-Current and Neutral-Current Neutrino-Nucleus
Scattering in a Relativistic Approach%
\thanks{Presented by Carlotta Giusti at the 45th Winter School in Theoretical Physics ``Neutrino Interactions: from Theory to Monte Carlo Simulations'', L\k{a}dek-Zdr\'oj, Poland, February 2--11, 2009.}%
}
\author{Carlotta Giusti, Andrea Meucci, Franco Davide Pacati
\address{Dipartimento di Fisica Nucleare e Teorica, Universit\`a degli Studi di 
Pavia\\
and Istituto Nazionale di Fisica Nucleare, Sezione di Pavia, Italy}
}
\maketitle
\begin{abstract}
Relativistic models developed for the exclusive and inclusive QuasiElastic (QE) 
electron scattering have been extended to Charged-Current (CC) and 
Neutral-Current (NC) $\nu$-nucleus scattering. The results of 
different descriptions of Final-State Interactions (FSI) are compared. 
\end{abstract}
\PACS{25.30.Pt, 13.15.+g, 24.10.Jv}

\section{Introduction}

Several decades of experimental and theoretical work on electron scattering have
provided a wealth of information on nuclear structure and dynamics \cite{book}. 
In these experiments the electron is the probe, 
whose properties are clearly specified, and the nucleus the target whose 
properties are under investigation. Additional information on
nuclear properties is available from  $\nu$-nucleus scattering.
Neutrinos can excite nuclear modes unaccessible in electron scattering, can
give information on the hadronic weak current and on the strange  form factors
of the nucleon. Although of great interest, such studies are not the only aim of 
many neutrino experiments, which are better devised for a precise determination 
of neutrino properties. In neutrino oscillation experiments nuclei
are used to detect neutrinos and a proper analysis of data requires that the  
nuclear response to neutrino interactions is well under control and that the  
unavoidable theoretical uncertainties on nuclear effects are reduced as much as possible.

In recent years different models developed and successfully tested in comparison
with electron scattering data have been extended to $\nu$-nucleus scattering. 
Although the two situations are different, 
electron scattering is the best available guide to determine the prediction power of a
nuclear model. 
Nonrelativistic and relativistic models have been developed to describe nuclear
effects with different approximations. They can be considered as alternative 
models, but only a relativistic approach is able to account for all the 
effects of relativity in a complete and consistent way. Relativity is important 
at all energies, in particular at high energies, and in the energy regime of 
many neutrino experiments a relativistic approach is required. 

Relativistic models for the exclusive and inclusive electron and neutrino
scattering in the QE region \cite{meucci1,ee,cc,nc} are presented in this 
contribution. In the QE region the nuclear response is dominated by one-nucleon 
knockout processes, where the probe interacts with a
quasifree nucleon that is emitted from the nucleus with a direct one-step
mechanism and the remaining nucleons are spectators.
In electron scattering experiments the outgoing nucleon can be detected in
coincidence with the scattered electron. In the exclusive $(e,e'p)$ reaction the
residual nucleus is left in a specific discrete eigenstate and the
final state is completely specified. In the inclusive $(e,e')$ scattering the 
outgoing nucleon is not detected and the cross section includes all the available
final nuclear states.

For an incident neutrino or antineutrino NC and CC scattering can be considered  
\begin{eqnarray}
\nu (\bar\nu) + A & \rightarrow & {\nu'} (\bar\nu') + N + 
( A - 1)  \hspace{3cm} \mathrm{NC} \nonumber \\
\nu (\bar\nu) + A & \rightarrow & l^{-} (l^{+}) +
p(n) + (A-1). \hspace{2.3cm} \ \mathrm{CC} 
 \nonumber\end{eqnarray} 
In NC scattering only the emitted nucleon can be detected and the cross 
section is integrated over the energy and angle of the final lepton. Also 
the state of the residual $(A-1)$-nucleus is not determined and the cross 
section is summed over all the available final states. The 
same situation occurs for the CC reaction if only the outgoing nucleon is 
detected. The cross sections are therefore semi-inclusive in the hadronic 
sector and inclusive in the leptonic one and can be treated as an $(e,e'p)$ 
reaction where only the outgoing proton is detected. The exclusive CC process 
where the charged final lepton is detected in 
coincidence with the emitted nucleon can be considered as well. 
The inclusive CC scattering where only the charged lepton is detected 
can be treated with the same models used for the inclusive $(e,e')$ reaction. 

For all these processes the cross section is obtained  in the one-boson 
exchange approximation from the contraction between the lepton tensor, that
depends  only on the lepton kinematics,  and the hadron tensor $W^{\mu\nu}$,
that contains the nuclear response and whose components are given by products 
of the matrix elements of the nuclear current  $J^{\mu}$ between the initial 
and final nuclear states, i.e.,
\begin{equation}
W_{\mu\nu} = \sum_f \, \langle \Psi_f\mid J^{\mu}(\q) \mid \Psi_i\rangle \, 
\langle \Psi_i \mid J^{\nu\dagger}(\q)\mid \Psi_f\rangle \, 
\delta(E_i+\omega-E_f),
\label{eq.wmn}
\end{equation}
where $\omega$ and $\q$ are the energy and momentum transfer, respectively.
Different but consistent models to calculate $W^{\mu\nu}$ in QE electron and
$\nu$-nucleus scattering are outlined in the next sections.

\section{Exclusive one-nucleon knockout}

Models based on the Relativistic Distorted-Wave Impulse Approximation (RDWIA) 
have been developed \cite{meucci1,Ud,Kel} to describe the exclusive reaction 
where the outgoing nucleon is detected in coincidence with the 
scattered lepton and the residual nucleus is left in a discrete eigenstate 
$n$. In RDWIA the amplitudes of Eq. \ref{eq.wmn} are obtained in a 
one-body representation as 
\begin{equation}
\langle\chi^{(-)}\mid  j^{\mu}(\q)\mid \varphi_n \rangle  \ ,
\label{eq.dko}
\end{equation}
where $\chi^{(-)}$ is the s.p. scattering state of the emitted 
nucleon, $\varphi_n$ the overlap between the ground state of the target 
and the final state $n$, i.e., a s.p. bound state, and 
$j^{\mu}$  the one-body nuclear current. In the model the s.p. bound and
scattering states are consistently derived as eigenfunctions of a Feshbach-type
optical potential \cite{book,meucci1}. Phenomenological ingredients are adopted
in the calculations. The bound states are Dirac-Hartree solutions 
of a Lagrangian, containing scalar and vector potentials, obtained in the 
framework of the relativistic mean-field theory  \cite{adfx}. The scattering
state is calculated solving the Dirac equation with relativistic 
energy-dependent complex optical potentials \cite{chc}.
RDWIA models have been quite successful in describing  a large amount of data 
for the exclusive  $(e,e^{\prime}p)$  reaction \cite{book,meucci1,Ud,Kel}. 

\section{Semi-inclusive neutrino-nucleus scattering}

The transition amplitudes of the NC and CC processes where only the outgoing 
nucleon is detected are described as the sum of the RDWIA amplitudes in 
Eq. \ref{eq.dko} over the states $n$. 
In the calculations \cite{nc} a pure Shell-Model (SM) description is assumed, i.e., 
$n$  is a one-hole state and the sum is over all the occupied SM states. FSI 
are described by a complex optical potential whose imaginary part reduces the 
cross section by $\sim 50\%$. A similar reduction is obtained in the RDWIA 
calculations for the exclusive one-nucleon knockout. 
The imaginary part accounts for the flux lost in a specific channel towards 
other channels.
This approach is conceptually correct for an exclusive reaction, where only 
one channel contributes, but it would be wrong for the inclusive scattering, 
where all the channels contribute and the total flux must be conserved. For the 
semi-inclusive process where an emitted nucleon is detected, some of the 
reaction channels which are responsible for the imaginary part of the potential 
are not included in the experimental cross section and, from this point of view, it is correct to include the absorptive imaginary part.
Numerical examples in different kinematics are given in \cite{nc}.

\section{Inclusive lepton-nucleus scattering}
In the inclusive scattering where only the outgoing lepton is
detected FSI are treated in the Green's Function Approach (GFA) 
\cite{ee,cc,eenr}. In this model the components of the hadron tensor are 
written in terms of the s.p. optical model Green's function. This is the result 
of suitable approximations, such as the assumption of a one-body current and 
subtler approximations related to the IA. 
The explicit calculation of the s.p. Green's function is avoided by its 
spectral representation, which is based on a biorthogonal expansion in terms of 
a non Hermitian optical potential $\cal H$ and of its Hermitian conjugate 
$\cal H^{\dagger}$. Calculations require matrix elements of 
the same type as the RDWIA ones in Eq. \ref{eq.dko}, but involve 
eigenfunctions of both $\cal H$ and $\cal H^{\dagger}$, where the different 
sign of the imaginary part gives in one case an absorption and in the other 
case a gain of flux. Thus, in the sum over $n$ the total flux is redistributed
and conserved.  
The GFA guarantees a consistent treatment of FSI in the exclusive and in 
the inclusive scattering and gives a good description of $(e,e')$ data \cite{ee}.

\begin{figure}
\begin{center}
\epsfig{file=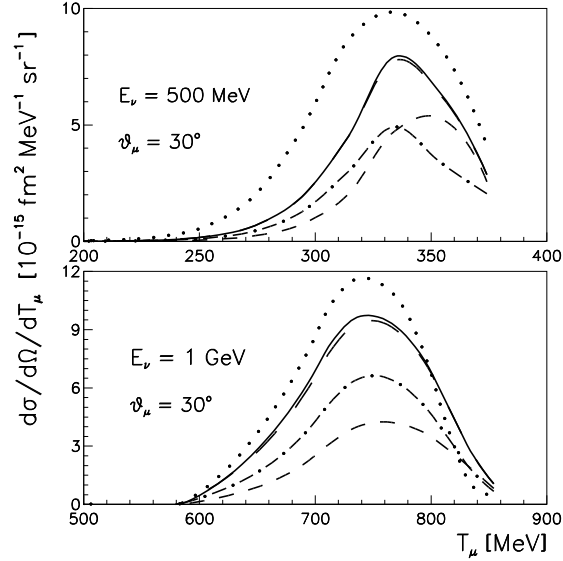,width=8cm}
\end{center}
\caption{The cross sections of the $^{16}O(\nu_{\mu},\mu^-)$ 
reaction for $E_\nu$ = 500 and 1000 MeV at $\theta_\mu = 30^{\mathrm o}$ as a
function of the muon kinetic energy $T_\mu$.  
Results for GFA (solid) RPWIA (dotted), rROP (long-dashed) are compared. 
The dot-dashed lines give the contribution of the integrated exclusive 
reactions with one-nucleon emission. Short dashed lines give the GFA 
results for the $^{16}O(\bar\nu_{\mu},\mu^+)$ reaction.
}
\label{fig1}
\end{figure}
An  example is displayed in Fig. \ref{fig1}, where the 
$^{16}O(\nu_{\mu},\mu^-)$  cross sections calculated in GFA are 
compared with the results of the Relativistic Plane Wave IA (RPWIA), where FSI 
are neglected. The cross sections obtained when only the real part of the 
Relativistic Optical Potential (rROP) is retained and the imaginary part is 
neglected are also shown in the figure. This approximation conserves the flux, 
but it is conceptually wrong because the optical potential has to be complex 
owing to the presence of inelastic channels. The partial 
contribution given by the sum of all the integrated exclusive one-nucleon 
knockout reactions, also shown in the figure, is much smaller than the 
complete result. The difference is due to the spurious loss of flux produced 
by the absorptive imaginary part of the optical potential. 

\begin{figure}
\begin{center}
\epsfig{file=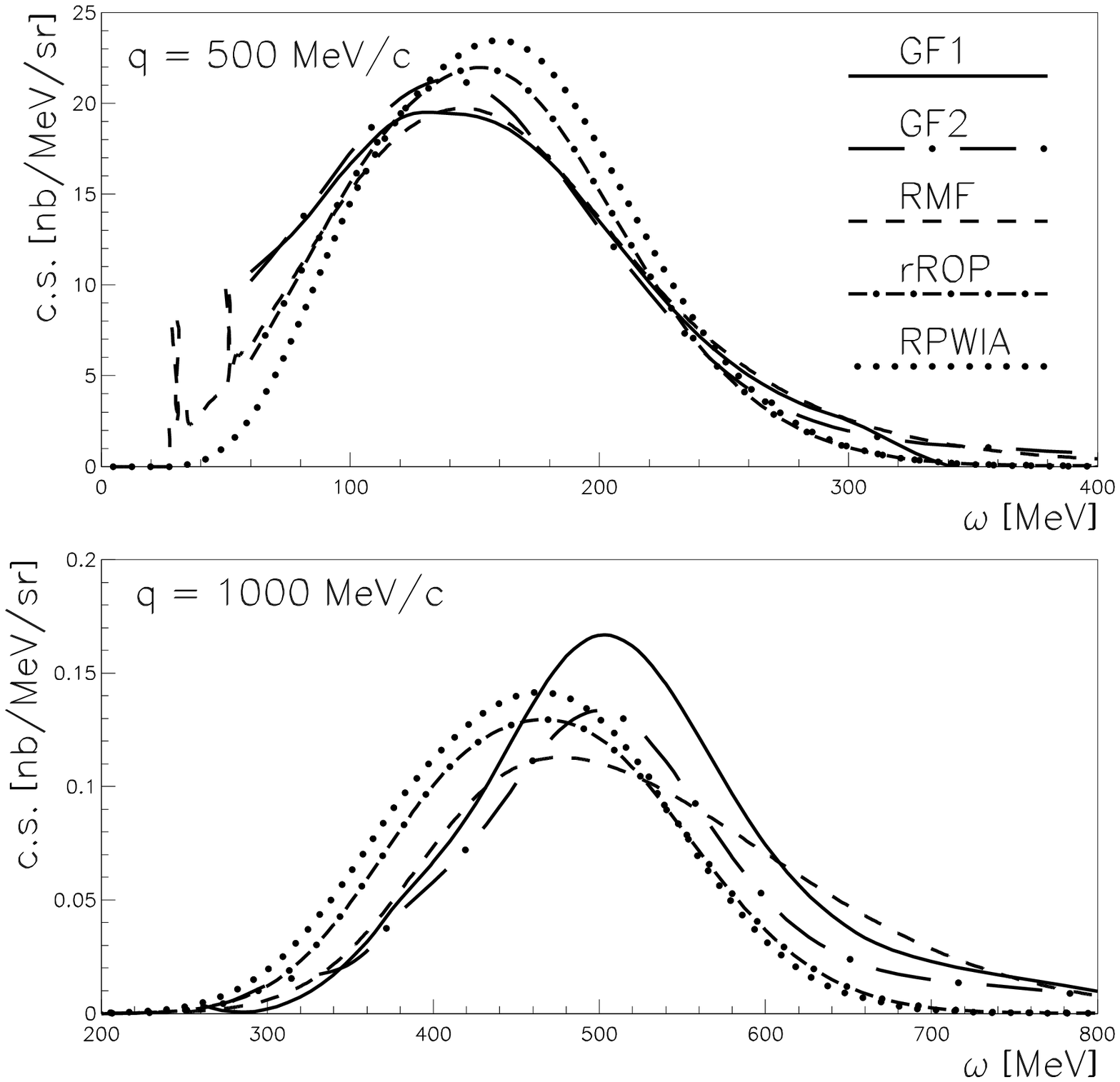,width=8cm}
\end{center}
\caption{The cross sections of the $^{12}C(e,e')$ 
reaction for an incident electron energy  of 1 GeV, $q$ = 500 (top panel) and
1000 MeV/$c$ (bottom panel), with RPWIA (dotted), 
rROP (dot-dashed), RMF (dashed), and  GFA with two optical potentials, EDAD1 
(GF1 solid) and EDA2 (GF2 long dot-dashed) \cite{chc}. 
}
\label{fig2}
\end{figure}
The analysis of data requires a precise knowledge of $\nu$-nucleus cross 
sections, where theoretical uncertainties on nuclear effects are reduced as much as possible. To 
this aim, it is important to check the consistency of different models and 
the validity of the approximations.   
The results of the relativistic models developed by our group and the 
Madrid-Sevilla group for the inclusive electron scattering are
compared in \cite{comp}.
An example is shown in  Fig. \ref{fig2} for the  $^{12}C(e,e')$ cross sections
calculated with different descriptions for FSI: 
RPWIA, rROP, GFA (with two parametrizations of the optical 
potential), and the Relativistic Mean Field (RMF) \cite{cab}, 
where the scattering wave functions are calculated with the same real 
potential used for the initial bound states.  
The differences between RMF and GFA increase with $q$: they are small 
at $q$ = 500 MeV$/c$ and significant at $q$ = 1000 MeV$/c$.
The  RMF cross section shows an asymmetry, with a long tail extending towards 
higher values of $\omega$. 
A less significant  asymmetry is obtained for both GFA cross sections, that  
at $q$ = 1000 MeV$/c$ are higher than the RMF one in the maximum region. 
The enhancement is different for the two optical potentials. 
The behaviour of the RMF and GFA results as a function of $q$ and
$\omega$  can be understood if we consider that RMF is based on the 
use of strong energy-independent scalar and vector real potentials, while GFA 
on a complex energy-dependent optical potential. Different values of $q$ and $\omega$ 
are sensitive to the behavior of the optical potential at different energies, 
and higher values correspond to higher energies. 
The GFA results are consistent with the general behavior of the 
optical potentials and are basically due to their imaginary part.  
Different parameterizations give similar real terms and the rROP cross sections 
are practically insensitive to the choice of optical potential. 
The real part decreases increasing the 
energy and the rROP result approaches the RPWIA one for large values of $\omega$. 
In contrast, the imaginary part has its maximum strength around 500 MeV and is
sensitive to the parameterization of the ROP. 
The imaginary part gives large differences between GFA and rROP in 
Fig. \ref{fig2}, while only negligible differences are obtained in the different
situation and kinematics of Fig. \ref{fig1}. 



\begin{thebibliography}{}

\bibitem{book}
S. Boffi, C. Giusti, F. D. Pacati, and M. Radici,
{\it Electromagnetic Response of Atomic Nuclei}, Oxford Studies in Nuclear
Physics, Vol. 20 (Clarendon Press, Oxford, 1996);
S. Boffi, C. Giusti, and F. D. Pacati, Phys. Rep. {\bf 226} 1 (1993).

\bibitem{meucci1}
A. Meucci, C. Giusti, and F.D. Pacati, 
Phys. Rev. C {\bf  64}  014604 (2001);
Phys. Rev. C  {\bf  64} 064615 (2001).

\bibitem{ee}
A. Meucci, F. Capuzzi, C. Giusti, and F.D. Pacati,
Phys. Rev. C {\bf 67}  054601 (2003);
Nucl. Phys. {\bf A756} 359 (2005).

\bibitem{cc}
A. Meucci, C. Giusti, and F.D. Pacati,
Nucl. Phys. {\bf A739} 277 (2004).

\bibitem{nc}
A. Meucci, C. Giusti, and F.D. Pacati,
Nucl. Phys. {\bf A744} 307 (2004);
Nucl. Phys. {\bf A773} 250 (2006);
Acta Physica Polonica {\bf B 37}, 2279  (2006).

\bibitem{Ud}
J.M. Ud\'{\i}as {\sl et al.},
Phys. Rev. C {\bf 48}, 2731 (1993);
Phys. Rev. C {\bf 51}, 3246 (1995);
Phys. Rev. C {\bf 64}, 024614 (2001).

\bibitem{Kel}
J.J. Kelly, 
Adv. Nucl. Phys. {\bf 23}, 75 (1996).

\bibitem{adfx}
W. P\"oschl, D. Vretenar, and P. Ring, 
Comput. Phys. Commun. {\bf 103}, 217 (1997);
G.A. Lalazissis, J. K\"onig, and P. Ring, 
Phys. Rev. C {\bf 55}, 540  (1997);
 M.M. Sharma, M.A. Nagarajan, and P. Ring
 Phys. Lett. {\bf B312}, 377 (1993).

\bibitem{chc} 
B.C. Clark,  in {\sl Proc. of the Workshop on Relativistic Dynamics and
   Quark-Nuclear Physics}, ed. by M.B. Johnson and A. Picklesimer 
   (John Wiley \& Sons, New York, 1986), p. 302;
   E.D. Cooper {\sl et al.}, Phys. Rev. C {\bf 47}, 297 (1993).

\bibitem{eenr}
F. Capuzzi, C. Giusti, F.D. Pacati, 
Nucl. Phys.  {\bf A524}, 681 (1991); 
F. Capuzzi, C. Giusti, F.D. Pacati, D.N. Kadrev, 
Annals Phys. {\bf{317}}, 492 (2005).

\bibitem{comp}
A. Meucci, J.A. Caballero, C. Giusti, F.D. Pacati, and J.M. Ud\'{\i}as,
arXiv:0906.2645

\bibitem{cab} 
J.A. Caballero {\sl et al.}, Phys. Rev. Lett. {\bf 95}, 252502 (2005). 

\end{thebibliography}
\end{document}